\documentclass[a4paper,11pt]{article}

\pdfoutput=1
\usepackage{color}
\usepackage{epsfig}
\usepackage{amsmath}
\usepackage{amssymb}
\usepackage{graphicx}
\usepackage{jheppub} 

\newcommand{\beq}{\begin{equation}}
\newcommand{\eeq}{\end{equation}}

\preprint{UUITP-35/17 \\
	\phantom{~} \hfill TCDMATH 17-21}

\title{\boldmath Konishi OPE coefficient at the five loop order}

\author[\dagger,\mathcal{x}]{Alessandro Georgoudis,}
\author[\mathcal{x}]{Vasco Gon\c calves}
\author[\dagger,\mathcal{z}]{and Raul Pereira}

\affiliation[\dagger]{Department of Physics and Astronomy, Uppsala University
Box 516, SE-751 20 Uppsala, Sweden}
\affiliation[\mathcal{x}]{ICTP South American Institute for Fundamental Research, IFT-UNESP,\\
S\~ao Paulo, SP Brazil 01440-070}
\affiliation[\mathcal{z}]{School of Mathematics and Hamilton Mathematics Institute, Trinity College Dublin, \\Dublin, Ireland}

\abstract{We use the method of asymptotic expansions to study the OPE limit of a four-point function of protected operators in $\mathcal{N}=4$ SYM. We use a new method for evaluating the resulting propagator-type integrals and then extract the OPE coefficient with Konishi at the five loop order. 
\newline
\newline 
\newline
\newline 
\newline
\newline 
\newline
\newline 
\newline
\newline 
\newline
\newline 
\centerline{{\it dedicated to the memory of } David Gon\c calves} }

\begin{document}

\maketitle
\flushbottom

\renewcommand{\thefootnote}{\arabic{footnote}}

\section{Introduction}\label{Intro}
Integrability of $\mathcal{N}=4$ SYM paves the way for finite coupling description of many of its observables. A paradigmatic example is the planar anomalous dimension of single-trace operators (see \cite{Gromov:2013pga} for the most efficient approach to the spectrum).  Over the last few years there have been finite coupling proposals for other observables, such as polygonal Wilson-loops \cite{Basso:2013vsa}, three-point functions \cite{short} and more recently $n$-point functions \cite{Thiago&Shota}. 

It was crutial, for all these examples, to compare the results of the integrability proposals with more conventional perturbative computations. For example, the integrability proposal for the spectrum problem had to be corrected to account for the mismatch between integrability computation and the perturbative one \cite{Fiamberti:2007rj,Fiamberti:2008sh,Bajnok:2008bm}. More recently, the integrability proposal for three-point functions (from here on called hexagon approach) has been checked by several weak coupling perturbative computations \cite{Eden,Eden:2015ija,3loops,Vasco,EdenPaul,Basso:2017muf}. The four-loop check was especially important as it confirmed the resolution of the double pole singularity that showed up for the first time at the four loop level in the hexagon approach.

The goal of this paper is to extend the perturbative computation of the OPE coefficient of two $20$' operators and the Konishi to five loops. From the integrability point of view, the motivation to go ahead with such computation is to check if the regularization of the double pole singularity continues to hold without any change at the five loop level \cite{Future}.

A five-loop computation using the more standard approach involving Feynman integrals seems to be a daunting task. The easiest way to extract the OPE coefficient is to consider a four-point function where all external operators are the chiral primaries in the $20$'. As these are in the same multiplet of the stress-energy tensor, it is possible to obtain the correlator to very high loop order \cite{Eden:2012tu,Bourjaily:2016evz}. In order to obtain the structure constant, it then suffices to consider the OPE decomposition in the coincidence limit. Since this four-point function is only known at the integrand level, one has to use the method of asymptotic expansions  in order to take the OPE limit. At this point, the OPE coefficient is expressed in terms of  many Feynman integrals with two external legs (also known as propagator integrals or p-integrals). By using integration by parts identities (IBPs), these integrals can be rewritten in terms of a small set of simpler integrals. However, at five loops this step turns out to be a bottleneck even using standard IBP codes such as LiteRed \cite{Lee:2012cn} and FIRE \cite{Smirnov:2014hma}. At last, one has to find the expansions of the master integrals, which are not known at five loops. We developed an efficient method of obtaining such expansions, whose detailed explanation we defer to a separate publication \cite{pintegrals}.

Thus, the plan of the paper will be to introduce, in section two, the four-point function of $20$' operators at five loop level and to explain how the method of asymptotic expansions works for this case. Then we will explain the details of using IBP relations to express every integral in terms of master integrals. After that, we will sketch the main idea of how the master integrals can be computed. This will be only a sketch since we will prepare a second paper just with the computation of these integrals since the master integrals are interesting for other computations in any perturbative QFT. Then we group  the main ingredients and present the result for the OPE coefficient.  There is one appendix with more details on the use IBPs since this turned out to be a non-trivial task at the five loop order as well as some comments on finding a basis of master integrals with uniform transcendentality at each order in $\epsilon$. 
\section{Four-point function and OPE limit}
A direct computation of the OPE coefficient is extremely complicated. One would have to determine not only the three-point function
		\begin{equation}\label{3ptft}
		\langle \mathcal O_1(x_1) \mathcal O_2(x_2) \mathcal O_3(x_3) \rangle = \frac{c_{123}}{(x_{12}^2)^{\Delta_1+\Delta_2-\Delta_3}(x_{13}^2)^{\Delta_1+\Delta_3-\Delta_2}(x_{23}^2)^{\Delta_2+\Delta_3-\Delta_1}} \,,
		\end{equation}
		but also all two-point functions $\langle\mathcal O_i \mathcal O_i \rangle$ so that the operators in \eqref{3ptft}  are normalized.
	
An alternative approach has been used in the past, and it takes advantage of the fact that the integrand of the four-point function of protected operators is known at very high loop order \cite{Eden:2012tu}. The contribution of each OPE channel in the correlator can be singled out by considering its OPE limit, which can be obtained by taking $x_{12}$ to $0$.

The main object of our study is then the four-point function of $\mathcal{O}_{20'}$ operators. These operators transform in a symmetric traceless representation, so it is useful to introduce a null polarization vector $y$ and denote the operators by $\mathcal O(x,y)~=~y_I \,y_J \;\mathrm{tr}(\phi^I \phi^J)$, where tracelessness is guaranteed by the condition $y^2=0$. Since they belong to the multiplet of the stress-energy tensor and Lagrangian density $\mathcal{L}$, the loop corrections to the correlator can be shown to take a factorized form
\begin{align}
\langle\mathcal{O}_{20'}\mathcal{O}_{20'}\mathcal{O}_{20'}\mathcal{O}_{20'} \rangle =F^{(0)}(x_i,y_i)+  \frac{2(N_c^2-1)}{(4\pi^2)^4}R(x_i,y_i)\sum_{n=1}^{\infty}\lambda^n\,F^{(n)}(x_i)\,,
\end{align}
where $\lambda$ is the 't Hooft coupling $\lambda = g^2N_c/(4\pi^2)$, and all dependence in the $SO(6)$ polarization vectors $y_i$ is contained in the prefactor
\begin{align}\label{Rfactor}
R(x_i,y_i)=& \frac{y_{12}^4 y_{34}^4}{x_{12}^2 x_{34}^2} + \frac{y_{13}^4 y_{24}^4}{x_{13}^2 x_{24}^2} + \frac{y_{14}^4 y_{23}^4}{x_{14}^2 x_{23}^2}+\frac{y_{12}^2 y_{23}^2 y_{34}^2 y_{41}^2}{x_{12}^2 x_{23}^2 x_{34}^2 x_{41}^2} \bigl( x_{13}^2 x_{24}^2 - x_{12}^2 x_{34}^2 - x_{14}^2 x_{23}^2\bigr) \notag \\
&+ \frac{y_{12}^2 y_{24}^2 y_{43}^2 y_{31}^2}{x_{12}^2 x_{24}^2 x_{43}^2 x_{31}^2} \bigl( x_{14}^2 x_{23}^2 - x_{12}^2 x_{34}^2 - x_{13}^2 x_{24
}^2\bigr) \notag \\
&  + \frac{y_{13}^2 y_{32}^2 y_{24}^2 y_{41}^2}{x_{13}^2 x_{32}^2 x_{24}^2 x_{41}^2} \bigl( x_{12}^2 x_{34}^2 - x_{13}^2 x_{24}^2 - x_{14}^2 x_{23
}^2\bigr).
\end{align}
The tree-level contribution $F^{(0)}(x_i,y_i)$ will not play any role in what follows, so we refer the reader to \cite{Grisha} for its definition. The functions $F^{(n)}(x_i)$ contain the loop corrections to the four-point function and are defined as
\begin{align}
F^{(n)}= \frac{x_{12}^2x_{13}^2x_{14}^2x_{23}^2x_{24}^2x_{34}^2}{n!(-4\pi^2)^n}\int d^dx_5\dots d^dx_{4+n}\,\frac{P^{(n)}(x_i)}{\prod_{1\le i <j\le 4+n}x_{ij}^2} \,,
\end{align}
where $P^{(n)}(x_i)$ is a homogeneous polynomial in $x_{ij}^2$ which is symmetric under $S_{4+n}$ permutations and has uniform weight $-(n-1)$ at each point. These properties of $P^{(n)}$ follow from the singularity structure of the four-point function and from the fact that the Lagrangian density sits on the same supermultiplet of the external operator $\mathcal{O}_{20'}$. The planar part of these polynomials is given up to five loops by \cite{Eden:2012tu}
\begin{align}\label{poly}
P^{(1)}=&1,\, \ \ \ P^{(2)} = \frac{1}{48}x_{12}^2x_{34}^2x_{56}^2+\text{$S_6$ perm}\,,\ \ P^{(3)} =\frac{1}{20}(x_{12}^2)^2x_{34}^2x_{45}^2x_{56}^2x_{67}^2x_{73}^2+\text{$S_7$ perm}\,,\nonumber\\
P^{(4)} =&\frac{1}{24} x_{12}^2 x_{13}^2 x_{16}^2 x_{23}^2 x_{25}^2 x_{34}^2 x_{45}^2 x_{46}^2 x_{56}^2 x_{78}^6+\frac{1}{8}x_{12}^2 x_{13}^2 x_{16}^2
   x_{24}^2 x_{27}^2 x_{34}^2 x_{38}^2 x_{45}^2 x_{56}^4 x_{78}^4
\nonumber\\
 & \ \ \  \ \ \  -\frac{1}{16} x_{12}^2 x_{15}^2 x_{18}^2 x_{23}^2 x_{26}^2 x_{34}^2
   x_{37}^2 x_{45}^2 x_{48}^2 x_{56}^2 x_{67}^2 x_{78}^2+\text{$S_8$ perm} \,,\nonumber\\
    P^{\text{(5)}}&= -\frac{1}{2} x_{13}^2 x_{16}^2 x_{18}^2 x_{19}^2 x_{24}^4 x_{26}^2 x_{29}^2 x_{37}^2 x_{38}^2 x_{39}^2 x_{47}^2 x_{48}^2
   x_{56}^2 x_{57}^2 x_{58}^2 x_{59}^2 x_{67}^2 \nonumber \\
 & +\frac{1}{4} x_{13}^2 x_{16}^2 x_{18}^2 x_{19}^2 x_{24}^4 x_{26}^2 x_{29}^2 x_{37}^4 x_{39}^2 x_{48}^4 x_{56}^2 x_{57}^2
   x_{58}^2 x_{59}^2 x_{67}^2 \nonumber \\
&+\frac{1}{4} x_{13}^4 x_{17}^2 x_{19}^2 x_{24}^2 x_{26}^2 x_{27}^2 x_{29}^2 x_{36}^2 x_{39}^2 x_{48}^6 x_{56}^2 x_{57}^2
   x_{58}^2 x_{59}^2 x_{67}^2\nonumber \\
&+\frac{1}{6} x_{13}^2 x_{16}^2 x_{19}^4 x_{24}^4 x_{28}^2 x_{29}^2 x_{37}^4 x_{38}^2 x_{46}^2 x_{47}^2 x_{56}^2 x_{57}^2
   x_{58}^2 x_{59}^2 x_{68}^2\nonumber\\
 &-\frac{1}{8} x_{13}^4 x_{16}^2 x_{18}^2 x_{24}^4 x_{28}^2 x_{29}^2 x_{37}^2 x_{39}^2 x_{46}^2 x_{47}^2 x_{56}^2 x_{57}^2
   x_{58}^2 x_{59}^2 x_{69}^2 x_{78}^2\nonumber \\
& +\frac{1}{28} x_{13}^2 x_{17}^2 x_{18}^2 x_{19}^2 x_{24}^8 x_{36}^2 x_{38}^2 x_{39}^2 x_{56}^2 x_{57}^2 x_{58}^2 x_{59}^2
   x_{67}^2 x_{69}^2 x_{78}^2 \nonumber \\
 &+\frac{1}{12} x_{13}^2 x_{16}^2 x_{17}^2 x_{19}^2 x_{26}^2 x_{27}^2 x_{28}^2 x_{29}^2 x_{35}^2 x_{38}^2 x_{39}^2 x_{45}^2
   x_{46}^2 x_{47}^2 x_{49}^2 x_{57}^2 x_{58}^2 x_{68}^2+\text{$S_9$ perm} \,.
\end{align}
By construction, the integrand of the four-point function has weight 4 in all integration variables, so each term in the polynomials \eqref{poly} leads to a conformal integral. They are both UV and IR finite and due to their conformal nature they can only depend non-trivially on the two cross ratios
\begin{align}
u=z \bar z=\frac{x_{12}^2x_{34}^2}{x_{13}^2x_{24}^2}\,, \qquad\qquad v=(1-z) (1-\bar z) = \frac{x_{14}^2x_{23}^2}{x_{13}^2x_{24}^2}\,,
\end{align}
where we define the complex variables $z$ and $\bar z$ for later use.
One should note that permutations inside $S_4 \times S_n$ have a simple action on the cross ratios, while the permutations in the quotient group $S_{4+n}/(S_4 \times S_n)$ can in principle produce inequivalent conformal integrals.

\subsection{Asymptotic expansions}
 
At the end of the day, the five-loop four-point function depends on 200 genuine five-loop conformal integrals. However, this number can be reduced to 141 independent integrals by using magic identities \cite{Drummond:2006rz}. The idea is that any subintegral of a conformal integral is conformal itself, and the cross ratios defined by its external points are invariant under the permutations $(12)(34)$, $(13)(24)$ and $(14)(23)$ of $S_4$. By performing such permutations at the level of the subintegral, one can find equivalences between conformal integrals that a priori look distinct. For example, they can be used to identify 14 integrals to be given by the five-loop ladder integral, which is known exactly  \cite{Usyukina:1993ch}
	\begin{equation}
	\phi^{(5)}\left( z, \bar{z} \right)= - \sum_{i=0}^{5} \frac{\left( 10-i \right)!}{5! ~i ! \left( 5-i \right) !} \frac{ \text{log}^i \left( z \bar{z} \right)}{z-\bar{z}} \left( \text{Li}_{10-i}\left( z \right)-\text{Li}_{10-i}\left( \bar{z} \right)  \right)\,,
	\end{equation}

We can use conformal symmetry to send $x_4$ to infinity, so that all conformal integrals appearing in the four-point function depend only on three external points. These integrals are not known, but all we need is their behaviour in the OPE limit $x_{12}\ll x_{13}$,  which can be obtained with the method of asymptotic expansions. In this way we can identify the leading order of the conformal integrals with combinations of simpler two-point integrals, {\em i.e. p-integrals}.

The main idea of the method is to consider regions where each integration variable is either of the order of $x_{2}$ or of the order of $x_{3}$ (we set $x_1$ to zero for simplicity). In practice, at five loops we have to consider $2^5$ regions, where for each the integrand can be simplified in a different manner depending on the scale of each integration variable.  For example, in the region where all integration variables obey $x_{i}^2 \gg x_{2}^2$ we have
\begin{align}
\frac{1}{x_{2i}^2} = \sum_{n=0}^{\infty} \frac{(2\,x_{2}\cdot x_{i}-x_{2}^2)^n}{(x_{i}^2)^{1+n}}\label{eq:eqthatsimplifies}\,.
\end{align}
Analogously, any propagator which includes two points of different scales will be decomposed in a similar manner, so that denominators depend solely on variables of a given region. In this way the two regions can only mix through numerators, where the variables of one appear as external vectors in the integrand of the other. As we perform tensor decomposition and write everything in terms of scalar integrals, the three-point integral is effectively decomposed into a sum of products of two-point integrals.

However this expansion has a finite radius of convergence which is set by the ball  $x_{2}^2<x_i^2$. By extending the integration regions to the whole space we make the integrals strictly divergent which we resolve by introducing dimensional regularization.

Let us note that the divergences cancel once we add all regions that contribute at a given order in the cross ratios. This has been verified both at three and four loops, at five loops we checked it for the ladder diagram and  for conformal integrals whose asymptotic expansions lead only to lower loop p-integrals.

At this point the conformal integrals are expressed in terms of propagator-type integrals, which are given in their most generic form by 
\begin{align}\label{eq:pIntegrals}
\int & \frac{\mathrm d^dx_5 \mathrm d^dx_6 \mathrm d^dx_7\mathrm  d^d  x_8\mathrm  d^dx_9}{(x_{15}^2)^{a_1}(x_{16}^2)^{a_2}(x_{17}^2)^{a_3}(x_{18}^2)^{a_4}(x_{19}^2)^{a_5}(x_{25}^2)^{a_6}(x_{26}^2)^{a_7}(x_{27}^2)^{a_8}(x_{28}^2)^{a_9}(x_{29}^2)^{a_{10}}(x_{56}^2)^{a_{11}}(x_{57}^2)^{a_{12}}}\times \nonumber\\
&\times \frac{1}{(x_{58}^2)^{a_{13}}(x_{59}^2)^{a_{14}}(x_{67}^2)^{a_{15}}(x_{68}^2)^{a_{16}}(x_{69}^2)^{a_{17}}(x_{78}^2)^{a_{18}} (x_{79}^2)^{a_{19}}(x_{89}^2)^{a_{20}}}\,,
\end{align}
with all exponents $a_i$  integer. A simple dimensional analysis is sufficient to show that only the first term in the sums coming from  (\ref{eq:eqthatsimplifies}) contributes at the leading order in $u$ and $(1-v)$ \cite{Vasco}.

\section{Evaluating propagator-type integrals}

In general one would use magic identities to reduce the number of integrals needed and also to choose the simplest possible form of the conformal integral. However, at five loops, it will be crucial to consider as many conformal integrals as possible, so we systematically used magic identities to create many equivalent integrals. After asymptotic expansions, those integrals are expressed in terms of 1721 five-loop p-integrals, and also some lower-loop ones. 
The integrals obtained are not all independent, as one can use Integration By Parts (IBP) identities to find relations between them and reduce to a small set of simpler master integrals. Such reductions can be performed with a combination of the public codes LiteRed \cite{Lee:2012cn} and FIRE \cite{Smirnov:2014hma}, but this can sometimes be a non-trivial task. For instance, some of the reductions took more than 256 GB of RAM and so we were forced to ignore conformal integrals in which those p-integrals appeared. 

\subsection{Master integrals} \label{bootstrap}

After the IBPs step  all conformal integrals are expressed in terms of master p-integrals. The last step is to obtain the expansions for these integrals, which is a new complication compared to the lower-loop case, where all masters were known.

The evaluation of master integrals is useful on its own since its applicability goes way outside the computation of OPE coefficients in $\mathcal{N}=4$ SYM. We will just sketch here the main ideas that we have used, deferring to a second publication a complete analysis of all master integrals at five loop level that were involved here \cite{pintegrals}. 

In order to obtain expansions for the master integrals, it is important to stress the differences between the original conformal integrals and the p-integrals obtained after asymptotic expansions. While the conformal integrals are UV and IR finite, and can only depend on the cross ratios $u$ and $v$, the propagator-type integrals are divergent and depend on the spurious scale $x_{13}$. The fact that both the divergence and the spurious scale cancel out upon summation of all regions in the asymptotic expansion introduces very strong constraints on the expansions of the master integrals. 
	In order to make this clearer, let us give a simple five-loop example. One of the conformal integrals obtained from the polynomial in \eqref{poly} is given by
	\begin{equation}\label{C1}
	C_1= \int \frac{\mathrm d^4 x_5 \ldots \mathrm d^4 x_9 \quad x_{13}^4 x_{24}^6 x_{35}^2}{x_{15}^2 x_{16}^2 x_{26}^2 x_{27}^2 x_{29}^2 x_{34}^2 x_{36}^2 x_{37}^2 x_{45}^2 x_{48}^2 x_{56}^2 x_{58}^2 x_{59}^2 x_{78}^2 x_{79}^2 x_{89}^2} \,.
	\end{equation}
	It turns out that in the OPE limit the only region that contributes at leading order in $u$ is the one where all integration variables are close to the points $x_1$ and $x_2$
	\begin{align}\label{C1exp}
	\left. C_1 \right|_{|x_{12}|\ll |x_{13}|} &\sim x_{13}^2 \int\frac{\mathrm d^4 x_5 \ldots \mathrm d^4 x_9}{x_{15}^2 x_{16}^2 x_{26}^2 x_{27}^2 x_{29}^2 x_{56}^2x_{58}^2 x_{59}^2 x_{78}^2 x_{79}^2 x_{89}^2} \nonumber\\
	&\sim u^{-11 + \frac{5d}{2}} (x_{13}^2)^{-10+ \frac{5d}{2}} P^{(5)}(d) \,.
	\end{align}
	In this way, the expansion of the conformal integral is given by a single five-loop p-integral, whose expansion is not known. At this point one would reduce it to master integrals, but that is not necessary for this particular example. The propagator-type integral diverges at most as $\epsilon^{-5}$, so its  expansion is in general given as
			\begin{equation}
			P^{(5)}(d=4-2 \epsilon) = \frac{c_{1}}{\epsilon^{5}}+ \frac{c_{2}}{\epsilon^{4}}+ \frac{c_{3}}{\epsilon^{3}}+ \frac{c_{4}}{\epsilon^{2}}+ \frac{c_{5}}{\epsilon}+ c_6 + \mathcal O(\epsilon) \,.
			\end{equation}
	If we use this to expand \eqref{C1exp}, we obtain an expression with poles in $\epsilon$ and powers of $\log(x_{13}^2)$. For the OPE limit of the  conformal integral \eqref{C1} to be finite, the coefficients in the expansion of the propagator-type integral must obey			
			\begin{equation}
			c_1 = c_2 = c_3 = c_4 = c_5 = 0 \,.
			\end{equation}
	At this point the coincidence limit of the conformal integral depends only on the finite part of $P^{(5)}$, which can be fixed by using the so called magic identities \cite{Drummond:2006rz}. The three-loop sub-integral of \eqref{C1} formed by the integration points $x_7$, $x_8$ and $x_9$ is itself a four-point conformal integral
	\begin{equation}
	\frac{1}{x_{27}^2 x_{29}^2 x_{37}^2 x_{48}^2  x_{58}^2 x_{59}^2  x_{78}^2 x_{79}^2 x_{89}^2} = \frac{1}{x_{25}^4 x_{34}^2} \Phi(\tilde u, \tilde v) \,.
	\end{equation}
	Since the cross-ratios $\tilde u$ and $\tilde v$ of the three-loop sub-integral are invariant under the exchange of points $x_{2}\leftrightarrow x_3,\, x_{4}\leftrightarrow x_5$, we have
	\begin{equation}
		\frac{1}{x_{27}^2 x_{37}^2 x_{39}^2  x_{48}^2 x_{49}^2 x_{58}^2 x_{78}^2 x_{79}^2 x_{89}^2} = \frac{1}{x_{34}^4 x_{25}^2} \Phi(\tilde u, \tilde v) \,.
		\end{equation}
	In this way we find a different conformal integral which is equivalent to \eqref{C1}
		\begin{equation}
		C_2= \int \frac{\mathrm d^4 x_5 \ldots \mathrm d^4 x_9 \quad x_{13}^4 x_{24}^6 x_{35}^2}{x_{15}^2 x_{16}^2 x_{25}^2 x_{26}^2 x_{27}^2 x_{36}^2 x_{37}^2 x_{39}^2 x_{45}^2 x_{48}^2 x_{49}^2 x_{56}^2 x_{58}^2 x_{78}^2 x_{79}^2 x_{89}^2} \,.
		\end{equation}
	Once again, there is only one region of the asymptotic expansions that contributes in the OPE limit $x_{12}\rightarrow 0$, but in this case it turns out to be a product of lower-loop two-point integrals
		\begin{align}
		\left. C_2 \right|_{|x_{12}|\ll |x_{13}|} &\sim x_{13}^4 \int \frac{\mathrm d^4 x_5 \mathrm d^4 x_6}{x_{15}^2 x_{16}^2 x_{25}^2 x_{26}^2 x_{56}^2} \int \frac{\mathrm d^4 x_7 \mathrm d^4 x_8 \mathrm d^4 x_9}{x_{17}^2 x_{18}^2 x_{36}^2 x_{37}^2 x_{39}^2 x_{78}^2 x_{79}^2 x_{89}^2} \nonumber\\
		& \sim u^{-5 + d} (x_{13}^2)^{-10+ \frac{5d}{2}} P^{(2)}(d) P^{(3)}(d) \,.
		\end{align}
	Since the expansion of these integrals is known, we are able to confirm that the spurious scale and divergence cancel out, but also obtain the conformal integral in the OPE limit
		\begin{equation}
		\left. C_2 \right|_{|x_{12}|\ll |x_{13}|} \sim \frac{120\; \zeta(3) \zeta(5)}{u}  \,.
		\end{equation}
		At the same time this fixes the finite term of the five-loop p-integral to be
			\begin{equation}
					c_6 = 120\; \zeta(3) \zeta(5)	\,.
			\end{equation}
		While in this simple example it was not necessary to evaluate the integral $C_1$, in most conformal integrals there is no magic identity which leads to an expression without five-loop p-integrals. In those cases, it is actually important to perform the reduction to master integrals, so that the constraints obtained from convergence, conformal symmetry and magic identites depend on the smallest possible set of variables.

This type of identities can be easily automatized providing a very large amount of constraints on the five-loop master p-integrals.
As an example of the power of this method, we checked that the data obtained was sufficient to match the expansions of p-integrals obtained using integrability methods in \cite{Caetano:2016ydc}.
Unfortunately, the constraints from conformal symmetry presented above could not determine all master integrals that contribute to the OPE coefficient of the Konishi operator. They have nonetheless enabled a huge simplification of the problem, as one is left with the evaluation of only $8$ single $\epsilon$ orders of $7$ master integrals out of the original $169$ masters.

The evaluation of the undetermined master integrals can be done using the code HyperInt \cite{Panzer:2014caa}. This package automatizes the computation of integrals that are linearly reducible. The main idea is to write the Feynman parametrization of an integral and then find an order of integration of the Feynman parameters such that at each integration step $k$ 
\begin{align}
f_{k} (\alpha_{k+1}; \ldots)= \int_0^{\infty} \mathrm d \alpha_k \; f_{k-1}(\alpha_{k}; \alpha_{k+1}, \ldots) \,,
\end{align}
the function $f_k$ can be written as an hyperlogarithm in the next integration variable $\alpha_{k+1}$\footnote{see \cite{Panzer:2014caa} for more details.}. 
An integral is linearly reducible is such an order exists, and luckily all master integrals left had such property, so the method was applicable to them.
The code works much more efficiently for convergent integrals, but some of the masters needed were divergent. A simple strategy to circumvent this problem is to find and evaluate convergent integrals in $d=4$ that include the required masters in their IBP reductions.

\section{From integrals to conformal data}
The Konishi operator is the lowest dimension non-protected operator that appears in the OPE expansion of two $20'$ operators
\begin{align}
\mathcal{O}_{20'}(x,y_2)\mathcal{O}_{20'}(0,y_1)\sim\,\big(\textrm{protected}\big)+ c_{20'20'\mathcal K}\frac{y_{12}^2}{(x^2)^{1-\gamma_{\mathcal{K}}/2}}\mathcal{K}(x)+\dots \,,
\end{align} 
where $\gamma_{\mathcal{K}}$ and $ c_{20'20'\mathcal K}$ are the anomalous dimension and OPE coefficient of the Konishi operator respectively. 

We are interested in the $20$ representation of the R-charge group $SU(4)$ of the Konishi operator. This can be obtained by an appropriate choice of the polarization vectors $y_i$ of the external operators
\begin{align}
\sum_{n\ge 0}\lambda^{n}F^{n}(x_i) \underbrace{\rightarrow}_{x_1\rightarrow x_2,\, x_3\rightarrow x_4} \frac{1}{6x_{13}^{4}}(c_{\mathcal{K}}^2(a)u^{\frac{\gamma_{\mathcal{K}}}{2}}-1)(1+O(u)+O(1-v))&\label{eq:OPE4pt}\,.
\end{align}
The left-hand side of the equation can be computed in the OPE limit using the methods explained above. In particular, we obtain
\begin{align}
x_{13}^4 \,F^{(5)}(x_i) \underbrace{\rightarrow}_{x_1\rightarrow x_2,\, x_3\rightarrow x_4} &
-8 (-7364 - 1812 \zeta_3 + 414 \zeta_3^2 - 2688 \zeta_5 - 
	864 \zeta_3 \zeta_5 - 3717 \zeta_7 - 5292 \zeta_9) \nonumber\\& -16 \log (u)\;
(2971 + 468 \zeta_3 + 27 \zeta_3^2 + 1080 \zeta_5 + 
	1260 \zeta_7) \nonumber\\&+
	96 \log^2 (u) \;(161 + 27 \zeta_3 + 45 \zeta_5)  -72 \log^3 (u) \;(37 + 6 \zeta_3)\nonumber\\&+ 252 \log^4 (u) -\frac{54}{5} \log^5 (u) \,,
\end{align}
while the lower loop $F^{(n)}$ were computed in \cite{Eden,Vasco,EdenPaul}. The appearance of $\log$ terms in $F^{(5)}$ can be easily understood from the right-hand side of (\ref{eq:OPE4pt}) as coming from the anomalous dimension of the exchanged operator. Thus, this procedure not only gives the OPE coefficient but also the anomalous dimension. Since the latter has been known for a long time we were able to use it as a check of the correctness of our result. All the lower loop conformal data can be obtained from the lower loop four-point function so we are able to extract the OPE coefficient of the Konishi operator at five loops
\begin{align}
(c_{20'20'\mathcal K}^{2})^{(5)}=-\, 64\, (7364 + 1812 \,\zeta_3 - 414\, \zeta_3^2 + 2688\,\zeta_5 + 
864 \,\zeta_3 \zeta_5 + 3717 \,\zeta_7 + 5292 \,\zeta_9) \,.
\end{align}
It is interesting to note that all terms with $\pi$ cancel out from the final result, which depends only on odd $\zeta$ values, just like at lower loops. It is also curious to observe that multiple zeta values are absent from the final result, even though they could have appeared starting at transcendentality 8. 

\section{Conclusions}
The description of three-point functions in $\mathcal{N}=4$ SYM using integrability is still very recent. The proposal has passed several checks and this work will provide another crucial test. Performing this five-loop computation within the Hexagon framework would be very important as it would show if the procedure for the resolution of the singularity that first appears at the four loop level continues to hold at higher loops. 

There are a couple of extensions to this paper that would be interesting to pursue. One of them would be to obtain all four-point functions of half-BPS operators in planar $\mathcal{N}=4$ SYM. One possible approach to this problem is to combine both the integrability methods and the Lagrangian insertion procedure \cite{Chicherin:2015edu}. Notice that in this setup the four-loop result is still unknown, so it would be important to start with that.

Another possible direction is to obtain three-point functions for operators with higher spin. It is trivial to apply the method of asymptotic expansions to expand the integrals further in the cross ratios. The only hurdle is to perform the  IBP reductions to express the result in terms of master integrals, but this may still be feasible for the first higher spin operators. 

Finally, let us note that it is in principle possible to extend this result to higher loops. The master integrals will not be known, but just like at five loops it might be possible to bootstrap them from convergence and symmetry considerations. The main obstacle with pursuing such a computation will rely on the ability to efficiently perform IBP reductions at such high loop order.
\section*{Acknowledgments}
We thank Roman Lee, Alexander Smirnov, Yang Zhang and Oliver Schnetz for useful discussion and help. We are also indebted to Gregor k\"alin for helping us with C++ and to Paul for careful reading of the manuscript.
We also would like to thank Erik Panzer for explaining in detail how the program HyperInt works and for evaluating one integral. 

The computations were performed on resources provided by the Swedish National Infrastructure for Computing (SNIC)  at Uppmax and HPC2N.
We would like to thank Lars Viklund at HPC2N and Linus Nilsson at Uppmax for their assistance during the computations and for allowing us to run jobs over the time limit. 
V.G. is funded by FAPESP grant 2015/14796-7 and CERN/FIS-NUC/0045/2015. The work of AG is supported by the Knut and Alice Wallenberg Foundation under grant \# 2015-0083. A.G  would like to thank FAPESP grant 2016/01343-7 for funding part of his visit to ICTP-SAIFR in March 2017 where part of this work was done. R.P.  was funded under VR2016-03503 and VR2012-3269 and he was also supported by SFI grant 15/CDA/3472.

\appendix
\section{Higher orders in expansion of master integrals}

The public code FIRE picks a given integral that has to be reduced and tries to express it in terms of simpler integrals. For example, an integral is simpler than another if it has a lower number of denominators \footnote{Other criteria include the comparison of the sum of powers of denominators and numerators}. Sometimes a full reduction consumes too many resources, so it is useful to first reduce to integrals of twelve denominators and then parallelize a full reduction of the resulting integrals. The trick described had to be used in the reduction of the following integral
\begin{equation}
I=\int\frac{\mathrm d^dx_5\dots \mathrm d^dx_9}{(2\pi)^{d/2}}\frac{x_{58}^2 x_{69}^2 x_{79}^2}{x_{17}^2 x_{18}^2 x_{19}^2 x_{25}^2 x_{26}^4 x_{28}^2 x_{29}^4 x_{56}^2 x_{57}^2 x_{59}^2 x_{67}^2 x_{68}^2 x_{78}^2 x_{89}^2}\,.
\end{equation}  

It turns out that the expression of  $I$ in terms of master integrals was, by far, the most complicated for all integrals analyzed. It is expressed in terms of $118$ master integrals, but the $d$-dependent rational factors multiplying each master introduced spurious poles in $d-4$ at a very high order. For example, the expansion of one of the relevant masters starts at transcendentality 6
\begin{equation}
M=\int \frac{\mathrm d^dx_5\dots \mathrm d^dx_9}{x_{8}^2 x_{9}^2 x_{15}^2 x_{16}^2 x_{17}^2 x_{58}^2 x_{59}^2 x_{67}^2 x_{69}^2 x_{78}^4} = \frac{-36\zeta_3^2}{\epsilon}+O(\epsilon) \,,
\end{equation}
but its coefficient in the reduction of $I$ has a pole of order 8, which means that the finite order of $I$ depends on the $\epsilon^8$ order of the master integral $M$, where terms of transcendentality 15 first appear
\begin{equation}
I= \left(\sum_{i=-8}^{\infty}a_i\epsilon^i \right)\,M +\dots
\end{equation}
The dots represent the contribution of the other master integrals to the reduction and the constants $a_i$ are given  by the IBP reduction.

This can become a serious bottleneck, since these high orders in $\epsilon$ cannot be obtained with HyperInt or the constraints imposed by conformal symmetry. Fortunately, we know that the finite order of $I$  has at most transcendentality $9$\footnote{This comes from analyzing the asymptotic expansions and the fact that the integrals are finite}, so it suffices to find the transcendentality $9$ part of each integral at any required order. 

The easiest way to achieve this is to find appropriate prefactors and combination of integrals that make each order in $\epsilon$ have homogeneous transcendentality. Let us illustrate this with a simple five-loop integral
\begin{align}
\int\frac{\mathrm d^dx_5\dots \mathrm d^dx_9}{(2\pi)^{d/2}}\frac{1}{x_8^2 x_9^2 (x_{17}^2)^2 x_{19}^2 (x_{58}^2)^2 x_{59}^2 (x_{67}^2)^2 x_{68}^2}=& -\frac{6 \zeta_3}{\epsilon^3} + \frac{48 \zeta_3 - \frac{\pi^4}{10}}{\epsilon^2} \nonumber\\
&+ \frac{\frac{4 \pi^4}{5}- 144 \zeta_3 -232 \zeta_5}{\epsilon} + \mathcal{O}(\epsilon^0)\,.
\end{align}
Since three of the integration points are connected only to two propagators each, there are three trivial integrations that can be performed using the formula
\begin{equation}
\int\frac{\mathrm d^dx_5}{\pi^{d/2}(x_{58}^2)^a (x_{59}^2)^b} = \frac{1}{\epsilon\, G(1,1)}\frac{G(a,b)}{(x_{89}^2)^{a+b-d/2}}\,,
\end{equation}
where the function $G$ is given by
\begin{equation}
G(a,b)= \frac{\Gamma(a+b-d/2)\Gamma(d/2-a)\Gamma(d/2-b)}{\Gamma(a)\Gamma(b)\Gamma(d-a-b)} \,.
\end{equation}
Our five-loop example effectively becomes a two-loop integral
\begin{align}
\frac{G^2(1,2)G(1+\epsilon,2)}{\epsilon^3 \,G(1,1)^3}\int\frac{\mathrm d^dx_8 \mathrm d^dx_9}{(2\pi)^{d/2}}\frac{1}{x_8^2 x_9^2x_{19}^2 (x_{89}^2)^{1+\epsilon}(x_{18}^2)^{1+2\epsilon}}. 
\end{align}
The two-loop integral with non-integer powers in the denominator has been studied in the literature \cite{Panzer:2013cha} and it turns out to have homogeneous transcendentality at each order in $\epsilon$ after dividing by $(1-2\epsilon)$ 
\begin{align}
\frac{1}{1-2\epsilon}\int\frac{\mathrm d^dx_8 \mathrm d^dx_9}{(2\pi)^{d/2}}&\frac{1}{x_8^2 x_9^2x_{19}^2 (x_{89}^2)^{1+\epsilon}(x_{18}^2)^{1+2\epsilon}}= 6\, \zeta_3+ \frac{\pi^4}{10}\, \epsilon+ 
	232\,\zeta_5  \,\epsilon^2  + \epsilon^3 \left(\frac{113\, \pi^6}{189} - 
	320 \,\zeta_3^2 \right) \nonumber\\
	&+ \epsilon^4 \left( -\frac{32\, \pi^4 \,\zeta_3 }{3}+ 7327 \,\zeta_7\right) + \epsilon^5 \left( \frac{
	24331 \,\pi^8}{10500} + \frac{1944 \,\zeta_{3,5}}{5} - 
	18220 \,\zeta_3 \,\zeta_5\right)\nonumber\\
	&+
	 \epsilon^6\left( -\frac{2930 \,\pi^6 \,\zeta_3}{63} +\frac{27916 \,\zeta_3^3}{3} - \frac{5041 \,\pi^4 \,\zeta_5}{15} + \frac{676106 \,\zeta_9}{3}\right).
\end{align}
The same strategy can be applied for all other masters in the reduction of $I$ which have easy integrations. For the more complicated cases where there are no easy integrations, one can try to find a prefactor directly for the master integral. We have started with the ansatz 
\begin{equation}
\frac{\text{polynomial in } \epsilon}{(1-2\epsilon)^4}
\end{equation}
and were able to find a numerator for all required integrals. In order to check the validity of the result, we found other integrals of homogeneous transcendentality and observed that the results were independent of the basis chosen.

Another nontrivial check we performed was that  the divergent part of  $I$ matched the expansion obtained from conformal symmetry with the methods of Section \ref{bootstrap}.

\bibliography{biblio}

\providecommand{\href}[2]{#2}\begingroup\raggedright\begin{thebibliography}{10}

\bibitem{Gromov:2013pga}
N.~Gromov, V.~Kazakov, S.~Leurent and D.~Volin, \emph{{Quantum Spectral Curve
  for Planar $\mathcal{N} =$ Super-Yang-Mills Theory}},
  \href{http://dx.doi.org/10.1103/PhysRevLett.112.011602}{\emph{Phys. Rev.
  Lett.} {\bf 112} (2014) 011602}, [\href{https://arxiv.org/abs/1305.1939}{{\tt
  1305.1939}}].

\bibitem{Basso:2013vsa}
B.~Basso, A.~Sever and P.~Vieira, \emph{{Spacetime and Flux Tube S-Matrices at
  Finite Coupling for N=4 Supersymmetric Yang-Mills Theory}},
  \href{http://dx.doi.org/10.1103/PhysRevLett.111.091602}{\emph{Phys. Rev.
  Lett.} {\bf 111} (2013) 091602}, [\href{https://arxiv.org/abs/1303.1396}{{\tt
  1303.1396}}].

\bibitem{short}
B.~Basso, S.~Komatsu and P.~Vieira, \emph{{Structure Constants and Integrable
  Bootstrap in Planar N=4 SYM Theory}},
  \href{https://arxiv.org/abs/1505.06745}{{\tt 1505.06745}}.

\bibitem{Thiago&Shota}
T.~Fleury and S.~Komatsu, \emph{{Hexagonalization of Correlation Functions}},
  \href{http://dx.doi.org/10.1007/JHEP01(2017)130}{\emph{JHEP} {\bf 01} (2017)
  130}, [\href{https://arxiv.org/abs/1611.05577}{{\tt 1611.05577}}].

\bibitem{Fiamberti:2007rj}
F.~Fiamberti, A.~Santambrogio, C.~Sieg and D.~Zanon, \emph{{Wrapping at four
  loops in N=4 SYM}},
  \href{http://dx.doi.org/10.1016/j.physletb.2008.06.061}{\emph{Phys. Lett.}
  {\bf B666} (2008) 100--105}, [\href{https://arxiv.org/abs/0712.3522}{{\tt
  0712.3522}}].

\bibitem{Fiamberti:2008sh}
F.~Fiamberti, A.~Santambrogio, C.~Sieg and D.~Zanon, \emph{{Anomalous dimension
  with wrapping at four loops in N=4 SYM}},
  \href{http://dx.doi.org/10.1016/j.nuclphysb.2008.07.014}{\emph{Nucl. Phys.}
  {\bf B805} (2008) 231--266}, [\href{https://arxiv.org/abs/0806.2095}{{\tt
  0806.2095}}].

\bibitem{Bajnok:2008bm}
Z.~Bajnok and R.~A. Janik, \emph{{Four-loop perturbative Konishi from strings
  and finite size effects for multiparticle states}},
  \href{http://dx.doi.org/10.1016/j.nuclphysb.2008.08.020}{\emph{Nucl. Phys.}
  {\bf B807} (2009) 625--650}, [\href{https://arxiv.org/abs/0807.0399}{{\tt
  0807.0399}}].

\bibitem{Eden}
B.~Eden, \emph{{Three-loop universal structure constants in N=4 susy Yang-Mills
  theory}},  \href{https://arxiv.org/abs/1207.3112}{{\tt 1207.3112}}.

\bibitem{Eden:2015ija}
B.~Eden and A.~Sfondrini, \emph{{Three-point functions in ${\cal N}=4$ SYM: the
  hexagon proposal at three loops}},
  \href{http://dx.doi.org/10.1007/JHEP02(2016)165}{\emph{JHEP} {\bf 02} (2016)
  165}, [\href{https://arxiv.org/abs/1510.01242}{{\tt 1510.01242}}].

\bibitem{3loops}
B.~Basso, V.~Goncalves, S.~Komatsu and P.~Vieira, \emph{{Gluing Hexagons at
  Three Loops}},
  \href{http://dx.doi.org/10.1016/j.nuclphysb.2016.04.020}{\emph{Nucl. Phys.}
  {\bf B907} (2016) 695--716}, [\href{https://arxiv.org/abs/1510.01683}{{\tt
  1510.01683}}].

\bibitem{Vasco}
V.~Goncalves, \emph{{Extracting OPE coefficient of Konishi at four loops}},
  \href{https://arxiv.org/abs/1607.02195}{{\tt 1607.02195}}.

\bibitem{EdenPaul}
B.~Eden and F.~Paul, \emph{{Half-BPS half-BPS twist two at four loops in N=4
  SYM}},  \href{https://arxiv.org/abs/1608.04222}{{\tt 1608.04222}}.

\bibitem{Basso:2017muf}
B.~Basso, V.~Goncalves and S.~Komatsu, \emph{{Structure constants at wrapping
  order}},  \href{https://arxiv.org/abs/1702.02154}{{\tt 1702.02154}}.

\bibitem{Future}
B.~Basso and R.~Pereira, \emph{{Hexagon at five loop}},
  \href{https://arxiv.org/abs/To appear}{{\tt To appear}}.

\bibitem{Eden:2012tu}
B.~Eden, P.~Heslop, G.~P. Korchemsky and E.~Sokatchev, \emph{{Constructing the
  correlation function of four stress-tensor multiplets and the four-particle
  amplitude in N=4 SYM}},
  \href{http://dx.doi.org/10.1016/j.nuclphysb.2012.04.013}{\emph{Nucl. Phys.}
  {\bf B862} (2012) 450--503}, [\href{https://arxiv.org/abs/1201.5329}{{\tt
  1201.5329}}].

\bibitem{Bourjaily:2016evz}
J.~L. Bourjaily, P.~Heslop and V.-V. Tran, \emph{{Amplitudes and Correlators to
  Ten Loops Using Simple, Graphical Bootstraps}},
  \href{http://dx.doi.org/10.1007/JHEP11(2016)125}{\emph{JHEP} {\bf 11} (2016)
  125}, [\href{https://arxiv.org/abs/1609.00007}{{\tt 1609.00007}}].

\bibitem{Lee:2012cn}
R.~N. Lee, \emph{{Presenting LiteRed: a tool for the Loop InTEgrals
  REDuction}},  \href{https://arxiv.org/abs/1212.2685}{{\tt 1212.2685}}.

\bibitem{Smirnov:2014hma}
A.~V. Smirnov, \emph{{FIRE5: a C++ implementation of Feynman Integral
  REduction}}, \href{http://dx.doi.org/10.1016/j.cpc.2014.11.024}{\emph{Comput.
  Phys. Commun.} {\bf 189} (2015) 182--191},
  [\href{https://arxiv.org/abs/1408.2372}{{\tt 1408.2372}}].

\bibitem{pintegrals}
A.~Georgoudis, V.~Goncalves, E.~Panzer and R.~Pereira, \emph{{All 5 loop planar
  propagator master integrals}},  \href{https://arxiv.org/abs/to appear}{{\tt
  to appear}}.

\bibitem{Grisha}
B.~Eden, P.~Heslop, G.~P. Korchemsky and E.~Sokatchev, \emph{{Hidden symmetry
  of four-point correlation functions and amplitudes in N=4 SYM}},
  \href{http://dx.doi.org/10.1016/j.nuclphysb.2012.04.007}{\emph{Nucl. Phys.}
  {\bf B862} (2012) 193--231}, [\href{https://arxiv.org/abs/1108.3557}{{\tt
  1108.3557}}].

\bibitem{Drummond:2006rz}
J.~M. Drummond, J.~Henn, V.~A. Smirnov and E.~Sokatchev, \emph{{Magic
  identities for conformal four-point integrals}},
  \href{http://dx.doi.org/10.1088/1126-6708/2007/01/064}{\emph{JHEP} {\bf 01}
  (2007) 064}, [\href{https://arxiv.org/abs/hep-th/0607160}{{\tt
  hep-th/0607160}}].

\bibitem{Usyukina:1993ch}
N.~I. Usyukina and A.~I. Davydychev, \emph{{Exact results for three and four
  point ladder diagrams with an arbitrary number of rungs}},
  \href{http://dx.doi.org/10.1016/0370-2693(93)91118-7}{\emph{Phys. Lett.} {\bf
  B305} (1993) 136--143}.

\bibitem{Caetano:2016ydc}
J.~Caetano, O.~Gurdogan and V.~Kazakov, \emph{{Chiral limit of N = 4 SYM and
  ABJM and integrable Feynman graphs}},
  \href{https://arxiv.org/abs/1612.05895}{{\tt 1612.05895}}.

\bibitem{Panzer:2014caa}
E.~Panzer, \emph{{Algorithms for the symbolic integration of hyperlogarithms
  with applications to Feynman integrals}},
  \href{http://dx.doi.org/10.1016/j.cpc.2014.10.019}{\emph{Comput. Phys.
  Commun.} {\bf 188} (2015) 148--166},
  [\href{https://arxiv.org/abs/1403.3385}{{\tt 1403.3385}}].

\bibitem{Chicherin:2015edu}
D.~Chicherin, J.~Drummond, P.~Heslop and E.~Sokatchev, \emph{{All three-loop
  four-point correlators of half-BPS operators in planar $ \mathcal{N} $ = 4
  SYM}}, \href{http://dx.doi.org/10.1007/JHEP08(2016)053}{\emph{JHEP} {\bf 08}
  (2016) 053}, [\href{https://arxiv.org/abs/1512.02926}{{\tt 1512.02926}}].

\bibitem{Panzer:2013cha}
E.~Panzer, \emph{{On the analytic computation of massless propagators in
  dimensional regularization}},
  \href{http://dx.doi.org/10.1016/j.nuclphysb.2013.05.025}{\emph{Nucl. Phys.}
  {\bf B874} (2013) 567--593}, [\href{https://arxiv.org/abs/1305.2161}{{\tt
  1305.2161}}].

\end{thebibliography}\endgroup
\bibliographystyle{JHEP}

\end{document}